\documentclass[aps,prb,two column,showpacs]{revtex4}
\usepackage{graphicx}

\begin{document}
\title{Superconducting phase transitions in frustrated Josephson-junction arrays on a dice lattice}
\author{In-Cheol Baek}
\author{Young-Je Yun}
\author{Mu-Yong Choi}
\affiliation{Department of Physics, Sungkyunkwan University, Suwon 440-746, Republic of Korea}

\begin{abstract}
Transport measurements are carried out on dice Josephson-junction arrays with the frustration index $f=1/3$ and
1/2 which possess, within the limit of the $XY$ model, an accidental degeneracy of the ground states as a
consequence of the formation of zero-energy domain walls. The measurements demonstrate that both the systems
undergo a phase transition to a superconducting vortex-ordered state at considerably high temperatures. The
experimental findings are in apparent contradiction with the theoretical expectation that frustration effects in
the $f=1/3$ system are particularly strong enough to suppress a vortex-ordering transition down to near zero
temperature. The data for $f=1/2$ are more consistent with theoretical evaluations. The agreement between the
experiments and the Monte Carlo simulations of a $XY$ model for $f=1/3$ suggests that the order-from-disorder
mechanism for the removal of an accidental degeneracy may still be effective in the $f=1/3$ system. The transport
data also reveal that the dice arrays with zero-energy domain walls experience a much slower critical relaxation
than other frustrated arrays only with finite-energy walls.
\end{abstract}


\maketitle

Vortex matters in Josephson-junction arrays (JJA's) and superconducting wire networks exposed to a magnetic field
have been extensively studied in recent years. \cite{R1} Studies of the systems of a variety of geometries have
been concentrated especially on the nature of ordering transitions and excitations in the ordered state. More
recently, special attention is given to highly frustrated systems which are unable to simultaneously satisfy the
competing interactions and consequently have the highly degenerate ground states. \cite{R2, R3, R4, R5, R6, R7,
R8, R9, R10, R11, R12} Understanding how a highly frustrated system orders or fails to order is one of most
challenging problems in current condensed matter physics.

The infinite degeneracy of the ground state has been suggested in the systems on the triangular, kagome,
honeycomb, or dice lattice with the formation of zero-energy domain walls in the ordered vortex state. \cite{R3,
R4, R7, R8, R11, R12} The common cause to lift the accidental degeneracy is thermal fluctuations in the
continuous phase variable, that is, spin waves. \cite{R3, R8, R11, R12} The difference in free energy of the
fluctuations forces many systems to select a particular vortex pattern at finite temperatures. This mechanism of
lifting an accidental degeneracy is often referred to as order-from-disorder.

Frustrated JJA's on the dice (or $T_{3}$) lattice have attracted recent attention because of the possibility that
the order-from-disorder mechanism for the removal of an accidental degeneracy may not be effective in the
systems. Theoretical evaluations \cite{R12} on the frustrated $XY$ model, which most closely represents a JJA in
a magnetic field, have demonstrated that for the frustration index (defined as the magnetic flux through a rhombus
tile in units of flux quantum) $f=1/3$, frustration effects on a dice array are particularly large and the system
fails to order down to zero temperatures. The stabilization of a specific vortex pattern was expected to appear at
extremely low temperatures only when one goes beyond the limits of the $XY$ model by including the magnetic
interactions of currents present in a real JJA. The predicted vortex-ordering transition temperature for $f=1/3$
is as low as 0.01 in units of $J/k_{B}$, where $J$ is a junction coupling strength and $k_{B}$ the Boltzmann
constant. However, it was also expected that the dynamically quenched relaxation of the vortex patterns may
prevent an observation of the vortex-pattern ordering in experiments. For the fully-frustrated dice array with
$f=1/2$, a commensurate vortex pattern was expected to be similarly stabilized at $T \lesssim 0.1 J/k_{B}$ by the
magnetic interactions of currents in a JJA rather than by the fluctuations. \cite{R12} It was argued for $f=1/2$
that the order-from-disorder mechanism becomes effective only at extremely low temperatures due to the reduced
contribution to free energy from the fluctuations as a consequence of a hidden gauge symmetry and thus the
magnetic interactions of current is a more important source for the ordering of vortices.

The results of theoretical evaluations contradict the Monte Carlo (MC) simulations \cite{R10} of a $XY$ model
exhibiting a periodic vortex pattern for $f=1/3$ at $T \lesssim 0.2 J/k_{B}$ as a consequence of a phase
transition. For $f=1/2$, the simulations present a Berezinskii-Kosterlitz-Thouless-like (BKT-like) transition to a
dynamic glass state at $T \sim 0.05 J/k_{B}$.

The Bitter decoration experiments \cite{R9} on superconducting wire networks have also revealed a periodic vortex
pattern in the $f=1/3$ case and no trace of regular vortex ordering for $f=1/2$. For superconducting wire
networks, both the magnitude and the phase of superconducting order parameter can vary near the temperatures of
interest. Hence, a quite different situation may develop in JJA's where the phase fluctuations play a more
important role.

This paper presents the results of transport measurements on dice JJA's with $f=1/3$ and 1/2. The results show a
substantial departure from the theoretical behavior for $f=1/3$. The agreement between the experiments and the
Monte Carlo simulations for $f=1/3$ suggests that the order-from-disorder mechanism may still be effective for
ordering in the $f=1/3$ system. The data for $f=1/2$ are more consistent with theoretical evaluations.

\begin{figure}
\includegraphics[width=1.0\linewidth]{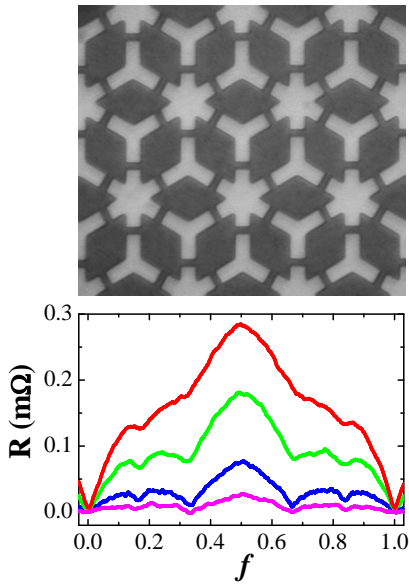}
\vskip 0.1true cm \caption{(top) Photograph of a dice array used in the experiment. (bottom) Sample resistance
with 30-$\mu$A excitation current plotted against frustration at different temperatures $T=4.4$, 4.25, 4.05, and
3.9 K in descending order.} \label{fig1}
\end{figure}

The experiments were performed on a dice array of 404$\times$525 Nb/Cu/Nb Josephson junctions. The photograph in
Fig.\ \ref{fig1} shows a portion of the JJA sample used in this work. Nb islands with a thickness of 0.2 $\mu$m
and an arm width of 4 $\mu$m were disposed on a 0.3-$\mu$m-thick Cu film periodically with the elementary side
length of the rhombus cell of 16 $\mu$m and a separation between adjacent islands of 1.4 $\mu$m. The variation of
the junction separation in the sample was less than 0.1 $\mu$m. Measurements of the resistance and the $IV$
characteristics were carried out by using the standard four-probe technique. The sample voltage was measured by a
transformer-coupled lock-in amplifier with a square-wave current at 23 Hz. The single-junction critical current
$i_c$ at low temperatures was obtained directly from the $IV$ curve. The $i_c$ and the junction coupling strength
$J (=\hbar i_{c}/2e)$ at high temperatures were determined by extrapolating the $i_c$ vs $T$ data at low
temperatures by the use of the de Gennes formula \cite{R13} for a proximity-coupled junction in the dirty limit.
Additional details of the measurements are described in Ref. 14.

\begin{figure}
\includegraphics[width=1.0\linewidth]{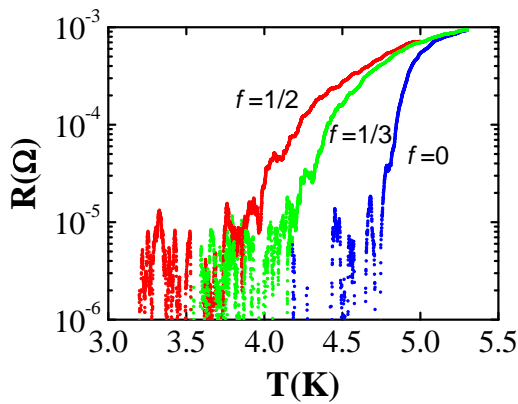}
\vskip 0.1true cm \caption{Temperature dependence of the sample resistance with 1-$\mu$A excitation current for
$f=0$, 1/3, and 1/2. The resistance is plotted in a log scale.} \label{fig2}
\end{figure}

Figure\ \ref{fig1} displays the resistance of the sample with 30-$\mu$A excitation current plotted against $f$ at
four different temperatures. Local minima are present at low order rationals $f=1/6$, 1/3, 2/3, and 5/6. A novel
feature appears at $f=1/2$. The anomalous positive cusp at $f=1/2$ does not disappear even at low temperatures
close to the superconducting transition temperature $T_{c} \sim 3.85$ K, determined from $IV$ characteristics
measurements (to be described below). The $f$-$R$ curve semi-quantitatively reflects the increase in the
mean-field superconducting transition temperature as a function of $f$. One can thus estimate from the curve the
energy cost for different vortex configurations. In general, the ground state at simple rational $f$ has a simple
periodicity. \cite{R15} The ground states for other $f$'s were found to consist of domains of nearby simple
rational $f$'s. \cite{R15, R16, R17} A finite energy required for exciting a domain wall causes a local minimum
of the system energy at the simple rational $f$'s. If the low $T_{c}$ and the formation of a dynamic vortex glass
proposed for $f=1/3$ in theory were the case, a positive cusp should appear at f=1/3. The appearance of a
distinct minimum at $f=1/3$ in the $f$-$R$ curve indicates, against the theoretical prediction \cite{R12}, that
the system with $f=1/3$ may have a vortex-ordered state with a locally minimal energy and a simple periodic
structure at considerably high temeperatures. The positive cusp at $f=1/2$ suggests that the system fails to order
within mean-field approximations and that excitations of domain walls cost an arbitrarily small energy. Apart
from the $f$-$R$ data, the $T$-$R$ data in Fig.\ \ref{fig2} of the sample with a smaller excitation current of 1
$\mu$A demonstrate that the $f=1/2$ system as well as the $f=1/3$ system may eventually make a superconducting
transition to an ordered state when the temperature is reduced below $\sim$3.9 K.

\begin{figure}
\includegraphics[width=1.0\linewidth]{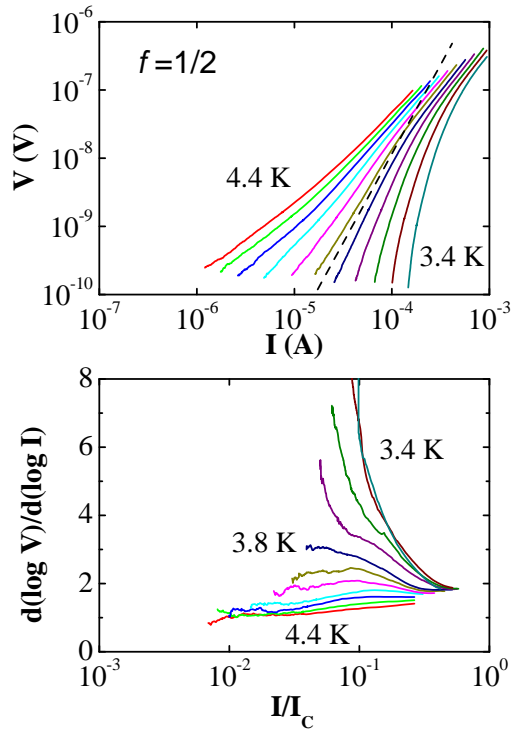}
\vskip 0.1true cm \caption{Evolution of $IV$ curves with temperature for $f=1/2$. The $I$ vs $V$ isotherms
differ by a temperature interval of 0.1 K. The panel at the bottom shows the slope of the isotherms as functions
of $I/I_{c}$. The dashed line is drawn to show where the superconducting phase transition occurs.} \label{fig3}
\end{figure}

The superconductivity for $f=1/2$ is verified by the $IV$ data plotted in Fig.\ \ref{fig3}. The distinct
activated character of the low-temperatures $IV$ curves reveals a superconducting state with a long-range phase
order as the ground state. As displayed more plainly in the $d({\log}V)/d({\log}I)$ vs $I/I_{c}$ plot of the same
data, the low-temperature activated character changes to the high-temperature resistive character at $T \sim
3.85$ K, corresponding to 0.10 in units of $J/k_{B}$. The $IV$ characteristics are quite similar to what observed
in frustrated square arrays experiencing a vortex-solid melting transition. \cite{R14, R17, R18, R19} The
similarity suggests that the superconducting-to-resistive transition at $T \sim 3.85$ K is a melting transition
of a vortex solid driven by domain walls. Even though the $f$-$R$ data are compatible with the MC simulations
\cite{R10} exhibiting the dynamic glass state at low temperatures, the $IV$ data appear to be more consistent
with the theoretical evaluations \cite{R12}. The positive cusp at $f=1/2$ can be related with the extreme
smallness of the fluctuation-induced free energy of zero-energy walls.

\begin{figure}
\includegraphics[width=1.0\linewidth]{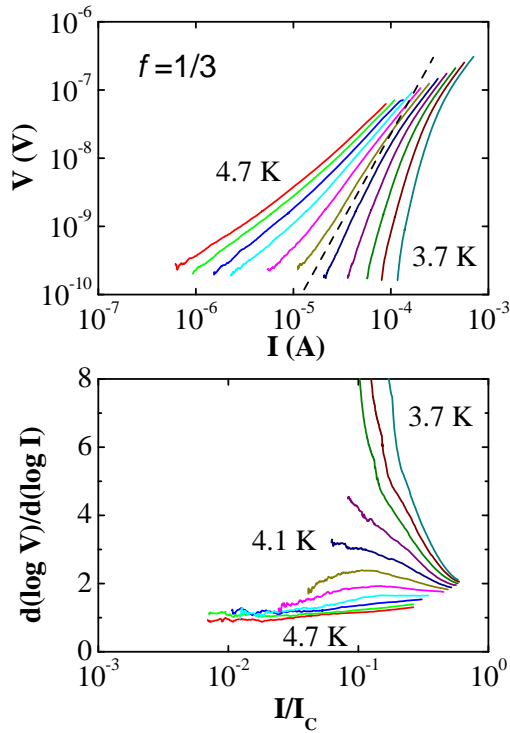}
\vskip 0.1true cm \caption{Evolution of the $IV$ curves with temperature for $f=1/3$.} \label{fig4}
\end{figure}

The $IV$ data for $f=1/3$ are shown in Fig.\ \ref{fig4}. The evolution of the $IV$ data with temperature is
similar to that for $f=1/2$. The $IV$ curves at low temperatures have an exponential form, indicative of a
vortex-ordered state with a long-range superconducting phase coherence. With the temperature raised, the
low-temperature activated behavior crosses over to a resistive form. The $IV$ data along with the $f$-$R$ data
indicate that the $f=1/3$ system undergoes a superconducting phase transition into an vortex-ordered state with a
simple periodicity at $T \sim 4.15$ K ($=0.2 J/k_{B}$). The $T_{c} \sim 0.2 J/k_{B}$ is comparable to those of
other frustrated arrays without any accidental degeneracy. \cite{R14, R17, R19} The occurrence of a
superconducting phase transition into a vortex-ordered state at such a high temperature is in apparent
contradiction with the theoretical argument \cite{R12} that frustration effects for $f=1/3$ are particularly
large enough to prevent the system from developing any order down to near zero temperature. However, it is in
agreement with the MC simulations \cite{R10} of a $XY$ model and compatible also with the decoration experiments
\cite{R9} on wire networks. The agreement between the experiments and the simulations may imply that the removal
of an accidental degeneracy for $f=1/3$ can be explained within the frame of a $XY$ model or that the
order-from-disorder mechanism may still be effective in the $f=1/3$ system.

Another interesting feature of our data is that the dynamic critical exponent $z$ of the systems determined from
the $IV$ data is unusually high. For a continuous superconducting transition, the $IV$ data satisfy a simple
power-law $IV$ relation, $V \sim I^{z+1}$, at $T_{c}$. \cite{R20} One can thus obtain $z$ directly from the
straight ${\log}I$ vs ${\log}V$ isotherm at $T_{c}$. $z$ of the dice arrays determined from the $IV$ data is
$\sim 1.7$ for both $f=1/3$ and 1/2, which is much larger than those ($z \lesssim 1$) of other frustrated arrays
\cite{R14, R17, R19} experiencing a vortex-lattice melting transition driven by finite-energy domain walls. The
exponential IV characteristics at $T < T_{c}$ do not allow one to relate the large $z$ to the possible occurrence
of a BKT-like transition. The large $z$ demonstrates that the critical relaxation in the dice arrays with an
accidental degeneracy is much slower than in other frustrated arrays without such an accidental degeneracy. It
may imply that the motion of zero-energy domain walls is not so effective in relaxing the system near $T_{c}$ as
that of finite-energy walls is.

To summarize, transport measurements were carried out on dice JJA's with $f=1/3$ and 1/2 which possess, within
the limit of the $XY$ model, an accidental degeneracy of the ground states as a consequence of the formation of
zero-energy domain walls. The measurements demonstrate that both the systems undergo a phase transition to a
superconducting vortex-ordered state at considerably high temperatures. The experimental findings are in apparent
contradiction with the theoretical expectation that frustration effects in the $f=1/3$ system are particularly
strong enough to suppress a vortex-ordering transition down to near zero temperature. The data for $f=1/2$ are
more consistent with theoretical evaluations. The agreement between the experiments and the MC simulations of a
$XY$ model for $f=1/3$ suggests that the order-from-disorder mechanism for the removal of an accidental
degeneracy may still be effective in the $f=1/3$ system. The transport data also reveal that the dice arrays with
zero-energy domain walls experience a much slower critical relaxation than other frustrated arrays only with
finite-energy walls.

\end{document}